\documentclass[prl,showpacs,preprintnumbers,amsmath,amssymb,twocolumn,superscriptaddress]{revtex4-1}
\usepackage{graphicx,pbox,lineno}
\usepackage[dvipsnames]{xcolor}
\usepackage[normalem]{ulem}

\begin{document}


\title{Two-band superconductivity with unconventional pairing symmetry in HfV$_2$Ga$_4$}

\author{A. Bhattacharyya}
\email{amitava.bhattacharyya@rkmvu.ac.in}
\affiliation{Department of Physics, Ramakrishna Mission Vivekananda Educational and Research Institute, Belur Math, Howrah 711202, West Bengal, India} 
\author{P. P. Ferreira}
\email{pedroferreira@usp.br}
\affiliation{Computational Materials Science Group (ComputEEL), Escola de Engenharia de Lorena, Universidade de S\~ao Paulo (EEL-USP), Materials Engineering Department (Demar), Lorena -- SP, Brazil}
\author{F. B. Santos}
\affiliation{Escola de Engenharia de Lorena, Universidade de S\~ao Paulo (EEL-USP), Materials Engineering Department (Demar), Lorena -- SP, Brazil}
\author{D. T. Adroja} 
\email{devashibhai.adroja@stfc.ac.uk}
\affiliation{ISIS Facility, Rutherford Appleton Laboratory, Chilton, Didcot, Oxon, OX11 0QX, United Kingdom} 
\affiliation{Highly Correlated Matter Research Group, Physics Department, University of Johannesburg, Auckland Park 2006, South Africa}
\author{J. S. Lord} 
\affiliation{ISIS Facility, Rutherford Appleton Laboratory, Chilton, Didcot, Oxon, OX11 0QX, United Kingdom} 
\author{L. E. Correa}
\affiliation{Escola de Engenharia de Lorena, Universidade de S\~ao Paulo (EEL-USP), Materials Engineering Department (Demar), Lorena -- SP, Brazil}
\author{A. J. S. Machado}
\affiliation{Escola de Engenharia de Lorena, Universidade de S\~ao Paulo (EEL-USP), Materials Engineering Department (Demar), Lorena -- SP, Brazil}
\author{A. L. R. Manesco}
\affiliation{Computational Materials Science Group (ComputEEL), Escola de Engenharia de Lorena, Universidade de S\~ao Paulo (EEL-USP), Materials Engineering Department (Demar), Lorena -- SP, Brazil}
\affiliation{Kavli Institute of Nanoscience, Delft University of Technology, Delft, The Netherlands}
\author{L T. F. Eleno}
\email{luizeleno@usp.br}
\affiliation{Computational Materials Science Group (ComputEEL), Escola de Engenharia de Lorena, Universidade de S\~ao Paulo (EEL-USP), Materials Engineering Department (Demar), Lorena -- SP, Brazil}

\begin{abstract}

In this letter, we have examined the superconducting ground state of the HfV$_2$Ga$_4$ compound using resistivity, magnetization, zero-field (ZF) and transverse-field (TF) muon-spin relaxation and rotation ($\mu$SR) measurements. Resistivity and magnetization unveil the onset of bulk superconductivity with $T_{\bf c}\sim$ 3.9~K, while TF-$\mu$SR measurements show that the temperature dependence of the superfluid density is well described by a nodal two-gap $s$+$d$-wave order parameter model. In addition, ZF muon relaxation rate increases with decreasing temperature below 4.6 K, indicating the presence of weak spin fluctuations. These observations suggest an unconventional multiband nature of the superconductivity possibly arising from the distinct $d$-bands of V and Hf ions with spin fluctuations playing an important role. To better understand these findings, we carry out first-principles electronic-structure calculations, further highlighting that the Fermi surface consists of multiple disconnected sheets with very different orbital weights and spin-orbit coupling, bridging the way for a nodal multiband superconductivity scenario. In this vein, therefore, HfV$_2$Ga$_4$-family stands out as an open avenue to novel unexplored unconventional superconducting compounds, such as ScV$_2$Ga$_4$ and ZrV$_2$Ga$_4$, and other many rare earths based materials.

\end{abstract}

\date{\today} 

\pacs{71.20.Be, 76.75.+i}

\maketitle


Unconventional superconducting materials have attracted considerable interest over the last decades due to their extremely rich physics and emerging breakthrough properties~\cite{scalapino2012, stewart2017}. In such materials, the superconducting state transcends the BCS-like signatures, as well as the isotropic $s$-wave pairing symmetry of the gap structure. Instead, a complex interaction framework among electrons, the crystal lattice, and spin-orbital fluctuations are established as the possible mediation mechanism of the Cooper pairs~\cite{norman2013}. HfV$_2$Ga$_4$ is a newly discovered superconducting compound with a critical temperature (T$_c$) of $\approx$\,3.9\, K that crystallizes in the tetragonal body-centered prototype YbMo$_2$Ga$_4$~\cite{Santos}. Substantial deviations of the temperature dependence of upper and lower critical fields from the expected Werthamer-Helfand-Hohenberg (WHH) formula~\cite{Werthamer} has led the authors to argue in favor of the presence of two superconducting gaps in the Fermi surface~\cite{Santos}. Later, first-principles electronic-structure calculations revealed the presence of electrons occupying very distinct bands at the Fermi level in the presence of spin-orbit coupling (SOC) effects~\cite{ferreira2018a}. Also, a substantial elastic anisotropy regime, due to the presence of extended linear vanadium chains in the structure~\cite{ferreira2018b}, indicates the feasibility of multiband superconductivity and the manifestation of possible unconventional properties.

\begin{figure*}[t]
\centering
\includegraphics[width=\linewidth, height=0.5\linewidth]{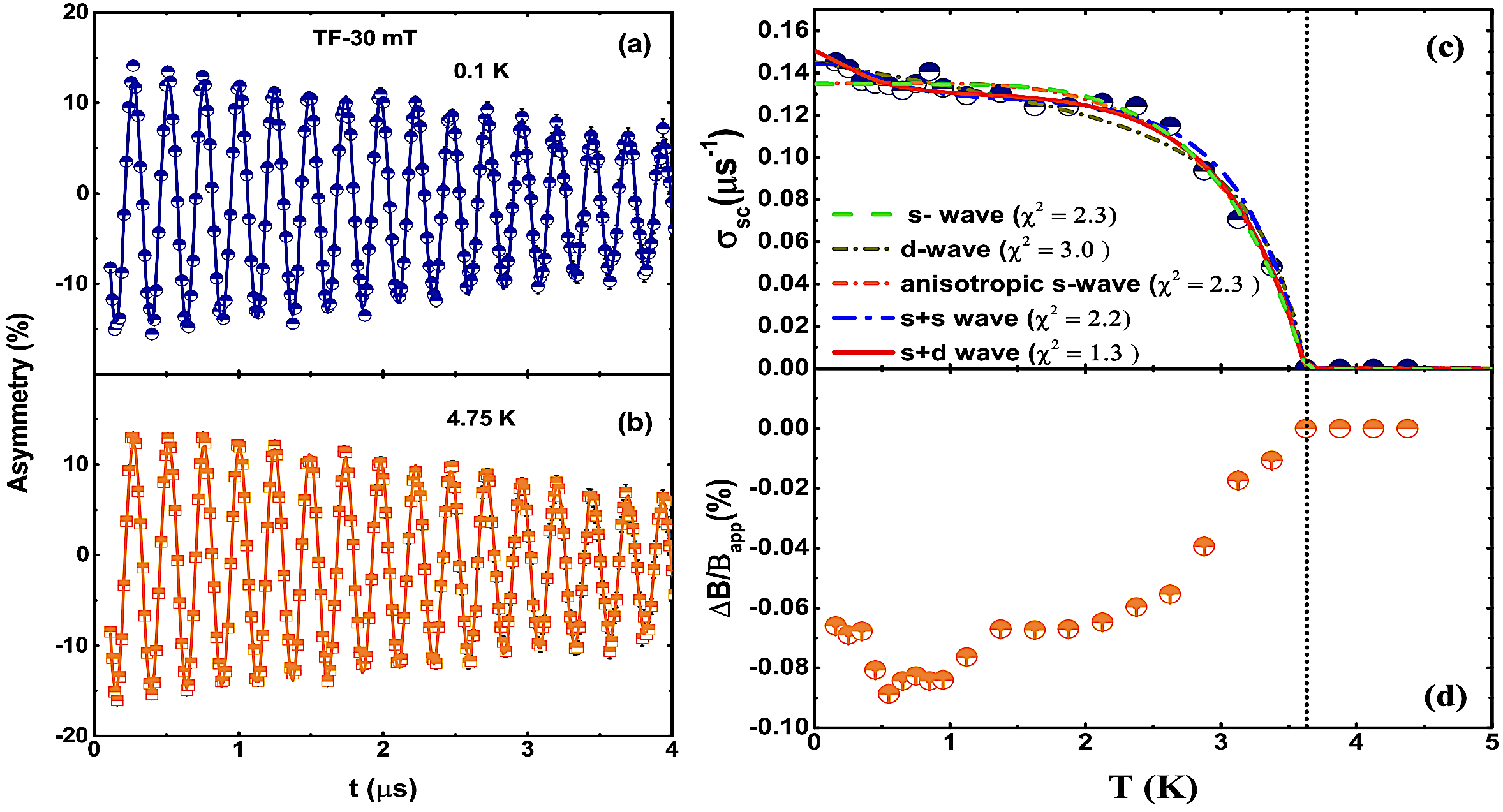}
\caption{Representative TF-$\mu$SR asymmetry spectra in the low time region collected at (a) at $T$ = 0.1 K and (b) at $T$ = 4.75 K (i.e. below and above $T_{\bf c}$) in an applied magnetic field of 30 mT. (c) Represent the $\sigma_{sc}(T)$ data in field cooling mode with fits using various gap models. The dash green line shows the fit using an isotropic single-gap $s$-wave model, the solid red line and dash-dotted blue line show the fit to a two-gap model, $s$+$d$-wave and $s$+$s$-wave, respectively. The orange dash-dotted line shows the fit using an anisotropic $s$-wave model and the dash-dotted purple line shows the fit using $d$-wave model. (d) Temperature dependence of the normalized internal field  from the sample.}
\label{tfmusr}
\end{figure*}

In this Letter, we present unambiguous evidence of two-gap nodal superconductivity in HfV$_2$Ga$_4$ using transverse (TF) muon spin rotation ($\mu$SR) measurement. However, in opposition to what is generally accepted in previous experimental and theoretical attempts, we have discovered an unconventional superconducting order parameter with $s$+$d$-wave pairing symmetry and spin fluctuations traces. These experimental findings are further supported by density functional theory (DFT) calculations. Therefore, HfV$_2$Ga$_4$ could represent a novel family of unconventional superconductors, going beyond Fe-based compounds, heavy-fermions, non-centrosymmetric systems, and other known classes~\cite{scalapino2012, stewart2017,norman2013}. Thus, several compounds within the HfV$_2$Ga$_4$-family, such as ScV$_2$Ga$_4$ and ZrV$_2$Ga$_4$ \cite{ferreira2018a, ferreira2018b}, stand out as an open avenue to further investigate unconventional superconducting properties using various experiments and theoretical models.  


For the present $\mu$SR study a high-quality polycrystalline sample of HfV$_2$Ga$_4$ was prepared by arc melting of the stoichiometric amount of hafnium, vanadium, and gallium on a water-cooled Cu crucible in a high-purity Ar atmosphere, the arc-melted pellet was encapsulated in an evacuated quartz ampoule and heated up to 800 $^{\circ}$C and kept at that temperature for one week, then quenched in cold water~\cite{Santos}. Magnetotransport measurements were performed using a VSM-PPMS EverCool II from Quantum Design. A powdered sample of HfV$_2$Ga$_4$ was used for the $\mu$SR experiments, which were carried out on the MUSR spectrometer at the ISIS Pulsed Neutron and Muon Source of Rutherford Appleton Laboratory, U.K. The sample was placed on a silver holder (99.999\%) using GE-varnish, which was loaded in a dilution refrigerator operating between 0.1 K and 4.75 K. Zero-field (ZF) and transverse field (TF)-$\mu$SR measurements were performed at different temperatures between 0.1~K and 4.75~K. For ZF-$\mu$SR measurement, an active compensation system was used to cancel any stray magnetic fields at the sample space to a level of $\sim 10^{-4}$~mT. ZF-$\mu$SR measurement is beneficial to identify the spontaneous internal field associated with time-reversal symmetry breaking~\cite{Sonier}. TF-$\mu$SR measurements were carried out in the presence of an external magnetic of 30 mT, which is well above the lower critical field ($\mu_{0}H_{c1}=1.2$ mT), and well below the upper critical field ($\mu_{0}H_{c2}=1.1$ T) of HfV$_2$Ga$_4$. The experimental data were analyzed using the WiMDA software~\cite{Pratt2000}.

\begin{figure*}[t]
\centering
\includegraphics[width=\linewidth, height=0.35\linewidth]{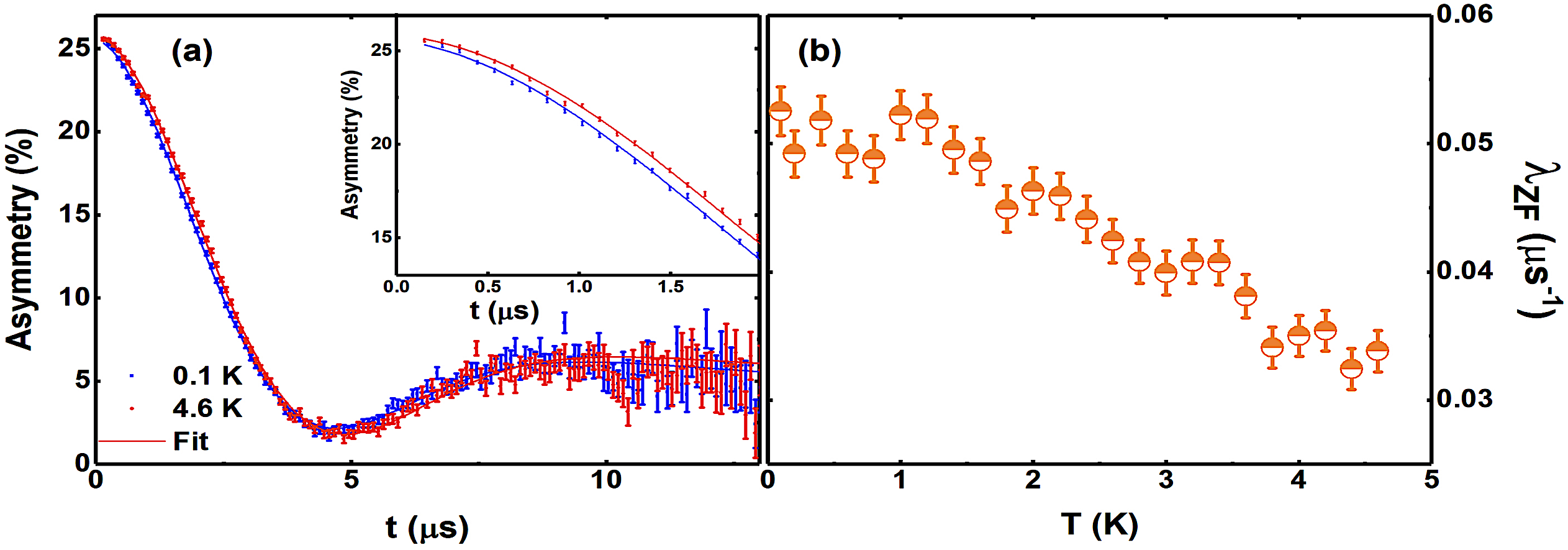}\hfill
\caption{(a) Represent the ZF-$\mu$SR spectra for HfV$_2$Ga$_4$ at 0.1 K and 4.6 K. The solid lines are fits to the data, as described in the text. The inset shows the asymmetry data at lower time. (b) shows the temperature dependence of the relaxation rate $\lambda_{ZF} (T)$, which intimate the presence of spin fluctuations.}
\label{zfmusr}
\end{figure*}



Santos {\it et al.}~\cite{Santos} recently reported superconductivity on HfV$_2$Ga$_4$ at $T_{\mathrm{C}}$ = 3.9 K, in which the upper and lower critical fields show unusual dependence with reduced temperature ($T/T_{\mathrm{C}}$), suggesting strong deviations from the conventional BCS behavior. This type of unusual behavior is also seen for other materials, such as Cu$_{0.3}$ZrTe$_{2-y}$~\cite{Machado1}, fluorine doped NdFeAsO~\cite{Adamski}, FeSe~\cite{Abdel-Hafiez}, SmFeAs$_{0.09}$F$_{0.01}$ , Ba$_{0.6}$K$_{0.4}$Fe$_{2}$As$_{2}$~\cite{Ren1,Ren2} and MgB$_{2}$~\cite{Li1}. To clarify the microscopic characteristics and superconducting gap structure of HfV$_2$Ga$_4$, TF-$\mu$SR measurements were carried in the mixed superconducting state. First, we applied an external magnetic field of 30 mT above $T_\mathrm{C}$, followed by colling down to 0.1 K. The TF-$\mu$SR asymmetry spectra were collected at various temperatures up to $T_{\mathrm{C}}$. Figs.~\ref{tfmusr}(a)-(b) exhibit two representative TF-$\mu$SR spectra collected at $T$ = 4.75 K and 0.1 K in $H$ = 30 mT. The asymmetry spectra at 0.1 K show faster relaxation compared to 4.75 K data as a result of the development of the flux line lattice below $T_{\mathrm{C}}$. The time evolution of the $\mu$SR asymmetry can be modeled by $A_\mathrm{TF}(t) = A_\mathrm{1}\cos(\gamma_\mathrm{\mu}B_1t+\Phi)\exp(-\frac{\sigma^{2}t^{2}}{2})+A_\mathrm{bg}\cos(\gamma_\mathrm{\nu}B_{bg}t+\Phi)$ \cite{Bhattacharyyarev, AdrojaThFeAsN, BhattacharyyaThCoC2,BhattacharyyaLu5Rh6Sn18}, where $A_1$ and $A_{bg}$ describe the initial asymmetries belonging to the sample and silver holder contributions, individually, with $A_{bg}$ not undergoing any depolarization; $B_1$ and $B_{bg}$ are the internal fields from the sample and from the sample holder, respectively. $\gamma_\mathrm{\nu}$/2$\pi$ = 135.53 MHz/T is the muon gyromagnetic ratio; $\Phi$ is the initial phase; and $\sigma$ is a Gaussian muon spin relaxation rate. The flux line lattice related muon relaxation can be extracted by subtracting the nuclear contribution according $\sigma_\mathrm{sc} = \sqrt{\sigma^{2}-\sigma_\mathrm{n}^2}$, where $\sigma_{\mathrm{n}}$ is the nuclear magnetic dipolar contribution which is temperature independent and was obtained from spectra measured above $T_\mathrm{C}$. By fitting the spectra at 4.75 K we obtained $\sigma_{\mathrm{n}}$ = 0.35 $\mu$s$^{-1}$. Considering that $\sigma_\mathrm{sc}$ is linked to the magnetic penetration depth ($\lambda$) by $\sigma_\mathrm{sc}~\approx~1/\lambda^2$, the superconducting gap symmetry can be determined from the temperature dependence of $\sigma_\mathrm{sc}(T)$. The temperature-dependent of magnetic penetration depth was analyzed employing different models, generally described by $\frac{\sigma_{sc}(T)}{\sigma_{sc}(0)} = \frac{\lambda^{-2}(T)}{\lambda^{-2}(0)} = 1 + \frac{1}{\pi}\int_{0}^{2\pi}\int_{\Delta(T,\phi)}^{\infty}\Big(\frac{\partial f}{\partial E}\Big) \frac{EdEd\phi}{\sqrt{E^{2}-\Delta(T,\phi)^2}}$\cite{Prozorov, AdrojaK2Cr3As3}, where $\Delta$ is an angle-dependent gap function, $f= [1+\exp(-E/k_\mathrm{B}T)]^{-1}$ is the Fermi function, and the integration signifies an average across the Fermi surface. The gap is expected to follow the function $\Delta(T) = \Delta_{0}\delta(T/T_\mathrm{C})\mathrm{g}(\phi)$, where $\Delta_0$ is the maximum gap value at zero temperature and $\mathrm{g}(\phi)$ is the angular dependence of the gap, equal to 1 and $\cos(2\phi)$ for an $s$- and $d$-wave model, respectively. Here $\phi$ is the azimuthal angle. The superconducting gap symmetry is expected to follow $\delta(T/T_\mathrm{C}) = \tanh[1.82[1.018(T_\mathrm{C}/T-1)]^{0.51}]$\cite{Pang2015, Annet1990}. This gap function is sufficiently precise to explain the temperature dependency at any coupling strength. 

The $\lambda^{-2}(T)$ data were fitted based on five different gap models (a) an isotropic $s$-wave gap, (b) an isotropic $s$+$s$-wave, (c) anisotropic $s$-wave, (d) a $d$-wave line node and (e) a nodal $s$+$d$-wave, as shown in Fig.~\ref{tfmusr}(c). The estimated fit parameters are given in the Supplemental Material~\cite{Supplemental} in Table I. Further, the diamagnetic signal observed below $T_{\mathrm{C}}$ can be seen through the decrease in the internal field below $T_{\mathrm{C}}$ as shown in Fig.~\ref{tfmusr}(d). From the fits presented in Fig.~\ref{tfmusr}(c) it is obvious that the isotropic $s$-wave, anisotropic $s$-wave, $d$-wave models do not fit the data as they give a high value of goodness of fit $\chi^2$. Contrariwise, two-gap models using the isotropic $s$+$s$-wave and a nodal $s$+$d$-wave show good fits to the $\lambda^{-2}(T)$ data. Furthermore the low $T$ upturn can be best fitted with a nodal $s$+$d$-wave model with a minimum value of $\chi^2$ = 1.3. The estimated parameters for the nodal $s$+$d$-wave model show one larger gap 2$\Delta_1(0)/k_BT_c$ = 6.27$\pm$0.2 (meV), which is larger than the value of 3.53 as expected for conventional BCS gap and another smaller gap 2$\Delta_2(0)/k_BT_c$ = 1.14$\pm$0.1 (meV). The smaller gap is a nodal gap for the $s$+$d$-wave model. Moreover, the large gap value indicates the presence of strong coupling superconductivity in HfV$_2$Ga$_4$. The parameters obtained from the fit to the $\sigma_{sc}(T)$ data of HfV$_2$Ga$_4$ using different gap models is presented in the Supplemental Material~\cite{Supplemental}. The multigap features are usually seen in iron-based superconductors, Ba$_{1-x}$K$_{x}$Fe$_{2}$As$_{2}$~\cite{twogaps, Khasanov2009}, cuprate superconductors~\cite{Khasanov2007, dwave} also in Bi$_{4}$O$_{4}$S$_{3}$~\cite{Biswas2013}. Following the method described in Ref.~\cite{Hillierns}, we determined the values of $\lambda_L(0)$= 797(4) nm for $s$+$d$ wave fit, $n_\mathrm{s} = 6.78(9) \times 10^{25}$ carriers m$^{-3}$, and $m^{*} = 1.528(2) m_\mathrm{e}$ respectively, for HfV$_2$Ga$_4$. For a detail calculations see the Supplemental Material~\cite{Supplemental}.

\begin{figure*}[t]
\centering
\includegraphics[width=0.6\linewidth, height=0.45\linewidth]{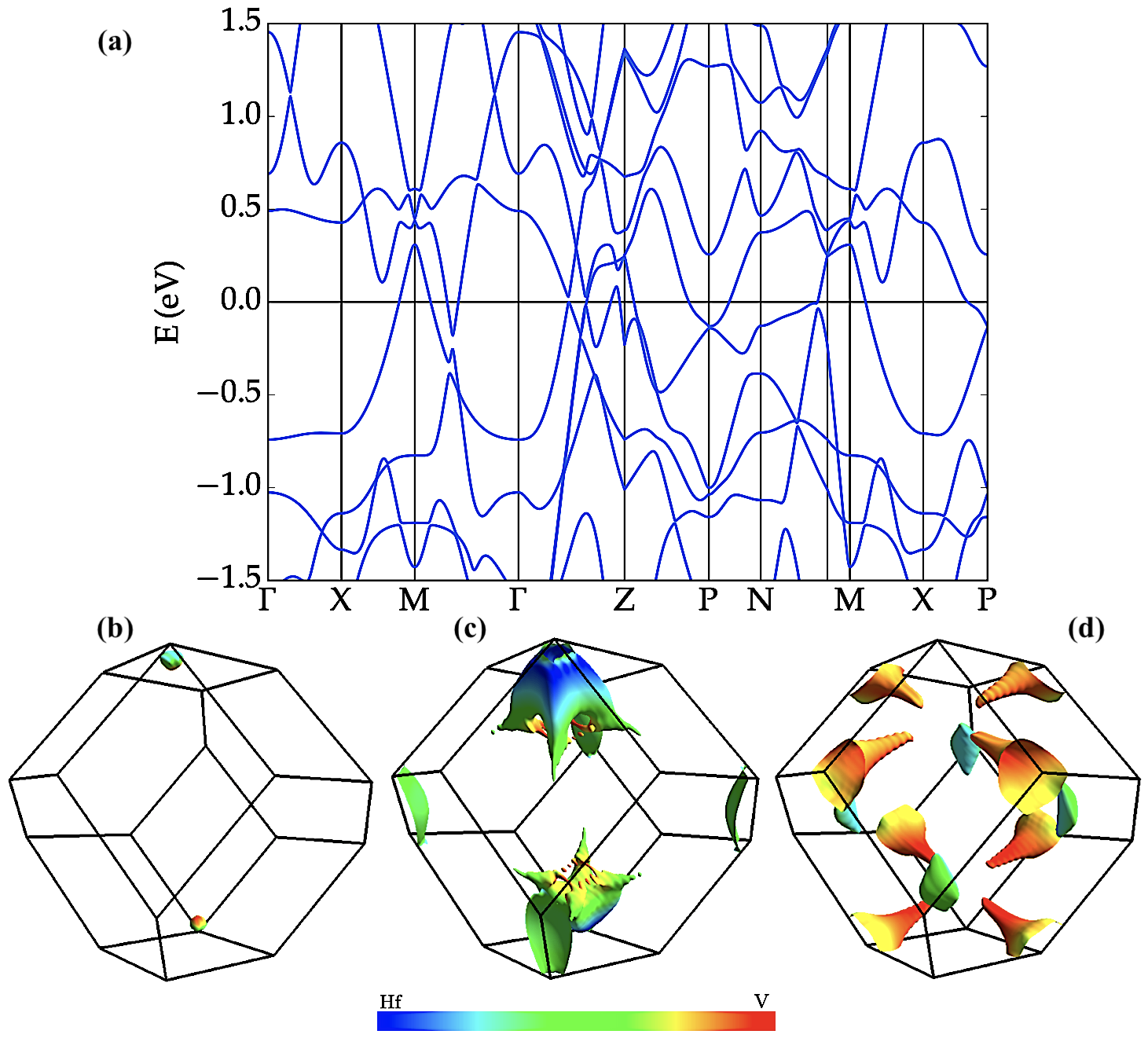}\hfill
\caption{(Upper panel) Electronic bands along a selected path in the first BZ of HfV$_{2}$Ga$_{4}$. The Fermi energy (E$_\mathrm{F}$) is set at 0.0 eV. (Lower panel) Hf- and V-d orbital character contribution to electronic states projected over the three distinct Fermi surface sheets.}
\label{fig:fermisurface}
\end{figure*}


To examine the fundamental issue of the presence of time-reversal symmetry (TRS) breaking or spin fluctuations in HfV$_2$Ga$_4$, we did ZF$-\mu$SR measurements. This technique is extremely helpful to identify the tiny spontaneous magnetic fields below $T_\mathrm{C}$. In the case of conventional superconductors, there is no change in the ZF muon relaxation rate ($\lambda_{ZF}$) below $T_\mathrm{C}$. $\lambda_{ZF}$ increases with decrease in temperature at $T_\mathrm{C}$ if TRS is broken. Fig.~\ref{zfmusr}(a) shows the ZF$-\mu$SR signal at 4.6 K and 0.1 K. The ZF$-\mu$SR signal could be best described by a combined Lorentzian and Gaussian Kubo-Toyabe relaxation function:~$A_\mathrm{ZF}(t) = A_\mathrm{2}A_\mathrm{KT}(t)e^{-\lambda_\mathrm{ZF}t}+A_\mathrm{bg}$, here $A_\mathrm{KT}(t) = [\frac{1}{3}+\frac{2}{3}(1-\sigma_\mathrm{KT}^{2}t^{2})\exp({-\frac{\sigma_\mathrm{KT}^2t^2}{2}})]$, is known as the Gaussian Kubo-Toyabe function, $A_\mathrm{2}$ and $A_\mathrm{bg}$ represent the asymmetry contribution from sample and silver holder, respectively. The resulting fit parameter is shown in Fig.~\ref{zfmusr}(b). It is interesting to note that $\lambda_{ZF}(T)$ increases with decreasing temperature, suggesting the presence of spin-fluatation in HfV$_2$Ga$_4$. Furthermore, the fits to the ZF data give $\sigma_\mathrm{KT} = 0.368(5) ~\mu \mathrm{s}^{-1}$ and $\lambda_\mathrm{\mu} = 0.0525(5) ~\mu \mathrm{s}^{-1}$ at 0.1~K and $\sigma_\mathrm{KT} = 0.368(5) ~\mu \mathrm{s}^{-1}$ and $\lambda_\mathrm{\mu} = 0.0337(6) ~\mu \mathrm{s}^{-1}$ at 4.6~K.


Analysis of band-structure and Fermi surface of the HfV$_2$Ga$_4$ compound can provide a suitable baseline from which to raise some phenomenological hypothesis of the mechanisms involved in such unconventional superconductivity shown by the experimental evidence presented in this letter. Therefore, optimized first-principles calculations were carried out in the framework of the Kohn-Sham scheme within the Density Functional Theory (DFT) \cite{hohenberg1964, kohn1965}, performed within the pseudopotential approach in the Perdew-Burke-Ernzerhof (PBE) generalized gradient approximation \cite{perdew1996, dal2014} as implemented in Quantum \textsc{Espresso} \cite{giannozzi2009, giannozzi2017} and support by auxiliary codes \cite{kawamura2019, kokalj1999}. 

Fig.~ref{fig:fermisurface}(a) shows the band-structure through high-symmetry points in the first Brillouin zone (BZ) \cite{setyawan2010}. Multiple distinct bands are crossing the Fermi energy ($E_F$), with very contrasting V-d and Hf-d characters. Other atomic orbital contributions to the electronic density in the vicinity of the $E_F$ are negligible. The obtained results are in great agreement with previous electronic-structure calculations \cite{ferreira2018a}, except for the presence of an additional band in the Fermi level (resulting in a very small quasi-spheric hole-pocket around the Z point in the Fermi surface). Additionally, Fig.~\ref{fig:fermisurface}(b-d) presents the projected-orbital Fermi surface with a color scheme for each irreducible representation. The reddish regions indicate a substantial contribution of V-d orbitals, which dictate the density of states (DOS) at the Fermi level; the bluish ones represent a major character derived from Hf-d orbitals; while the greenish branches depict a strong hybridization between these two states, as indicated by the color bar scale. 

The Fermi surface consists of multiple tridimensional disconnected sheets, bridging the way for a multiband superconductivity scenario: (a) a minor hybridized hole sphere next to the Z point; (b) a quasi-hyperbolic paraboloid with hole carriers that develops around the M point with substantial admixture of Hf-d and V-d orbitals, in the midpoint of the line that connects the two rhombuses (which we will call \emph{horse-saddle}), plus a very complex hole-like surface around the Z point composed mainly by Hf-d character with a slight degree of V-d (\emph{jellyfish}); (c) and multiple V-d electron cone-shaped pockets around the P point (\emph{seashell}), together with small electron-pockets saddled within the BZ along the M-$\Gamma$ high-symmetry line. 

Such hole bands with a predominant Hf-derived character are deeply sensitive against spin-orbit interactions, as opposed to V-d bands, which, in turn, play a crucial role in the low-energy electronic states \cite{ferreira2018a}. This fermiology, with very distinct disconnected pockets, in the sense of their orbital weight contribution and, hence, their effective spin-orbit coupling, favors the condensation of pairs with different superconducting order parameters. Since \emph{jellyfish}-pockets have a strong Hf-d character, we could expect the emergence of pair states with non-zero angular momentum and strong angular anisotropy, resulting in an even-parity momentum-dependent order parameter. On the other hand, the negligible effect of SOC in V bands favors the emergence of conventional $s$-wave pairing. This interpretation supports the experimental $s+d$-wave symmetry as founded. Moreover, the weighting factor of 0.57 to $s$-wave pairing is attributed to the higher contribution of V states on the DOS at the Fermi level, residing mainly at the \emph{seashell}-pockets, and a lower $d$-wave character coming from those Hf-derived structures.

On the other hand, repulsive interactions between the sheets, driven by spin fluctuations, for instance, are unexpected due to the complex nature of the fermiology. However, the \emph{jellyfish}- and \emph{horse-saddle}-pockets will take advantage of hole doping, increasing the SOC effects on the low-energy states, at the same time that \emph{seashell}-pockets will gradually decrease (see the Supplemental Material \cite{Supplemental}), paving the way for discontinuous sign change of the order parameter phase and spin fluctuations by the enlargement of possible nearly induced nesting instabilities \cite{kuroki2001, mazin2008, kuroki2008, singh2012}. This mechanism becomes relevant in our context since Ga-deficient HfV$_2$Ga$_4$ samples can, in a first approximation, effectively decrease the chemical potential of occupied states by a few meV, promoting such features. Also, hole doping promotes an increase in the density of states, favoring electronic correlations, and contributing to spontaneous symmetry breaking mechanisms. Indeed, nearly nested Fermi surface pockets and the anisotropic gap function support the enhanced pairing strength observed in $\mu$SR measurements (2$\Delta_1$/k$_B$T$_c$ = 6.27) \cite{terashima2009, monthoux1994a, pao1994}. Therefore, Ga-deficiency can be imperative to the unconventional superconductivity of HfV$_2$Ga$_4$, which seems to be unique, and high-quality single crystals, with systematic doping content, are urgently required to investigate the superconducting state further and to confirm the proposed hypothesis.


In summary, we have examined the superconducting state of HfV$_2$Ga$_4$ utilizing magnetotransport, $\mu$SR, and numerical band-structure calculations. The temperature dependence of the magnetic penetration depth, $\lambda^{-2}(T)$, is better fitted to a nodal two-gap $s$+$d$-wave model than a single gap isotropic $s-$wave, anisotropic $s$-wave or $d$-wave models, suggesting unconventional superconductivity in HfV$_2$Ga$_4$. The large value of gap to T$_{\mathrm{c}}$ ratio, 2$\Delta_1(0)/k_{\mathrm{B}}T_{\mathrm{c}}$ = 6.27$\pm0.2$, obtained from the nodal $s$+$d$-wave gap, is larger than 3.53, expected for conventional BCS superconductors, indicating the presence of strong coupling superconductivity in HfV$_2$Ga$_4$. The decrease in ZF relaxation rate with temperature indicates the presence of spin fluctuations in the superconducting state of HfV$_2$Ga$_4$. In addition, \emph{ab initio} calculations suggest that there are electrons derived from multiple distinct bands in disconnected sheets of the Fermi surface, in agreement with the experimental evidence of two-gap superconductivity for HfV$_2$Ga$_4$, and the momentum-dependent SOC interactions may be a theoretical starting point to elucidate the appearance of nodal superconductivity.


We want to acknowledge Mr. Kartik Panda for helping MUSR data analysis and Prof. A. M. Strydom for interesting discussions. AB would like to acknowledge the Department of Science and Technology (DST) India, for an Inspire Faculty Research Grant (DST/INSPIRE/04/2015/000169), and the UK-India Newton grant for funding support. DTA would like to thank the Royal Society of London for the UK-China Newton funding and the Japan Society for the Promotion of Science for an invitation fellowship. FBS, PPF, LEC, AJSM, ALRM, and LTFE gratefully acknowledge the financial support of the Brazilian agencies Conselho Nacional de Desenvolvimento Cientifico e Tecnologico (CNPq) under Grants No. 302149/2017-1 and 431868/2018-2, and Fundac\~ao de Amparo \`a Pesquisa do Estado de S\~ao Paulo (FAPESP) under Grants No. 2016/11774-5, 2016/11565-7, 2016/10167-8, 2018/08819-2, 2018/10835-6, 2018/20546-1, 2019/05005-7, and 2019/07082-9.

\newpage
\section{Supplemental Material:~Two-band superconductivity with unconventional pairing symmetry in HfV$_2$Ga$_4$}

\begin{figure*}[htbp]
\centering
\includegraphics[width=\linewidth, height=0.3\linewidth]{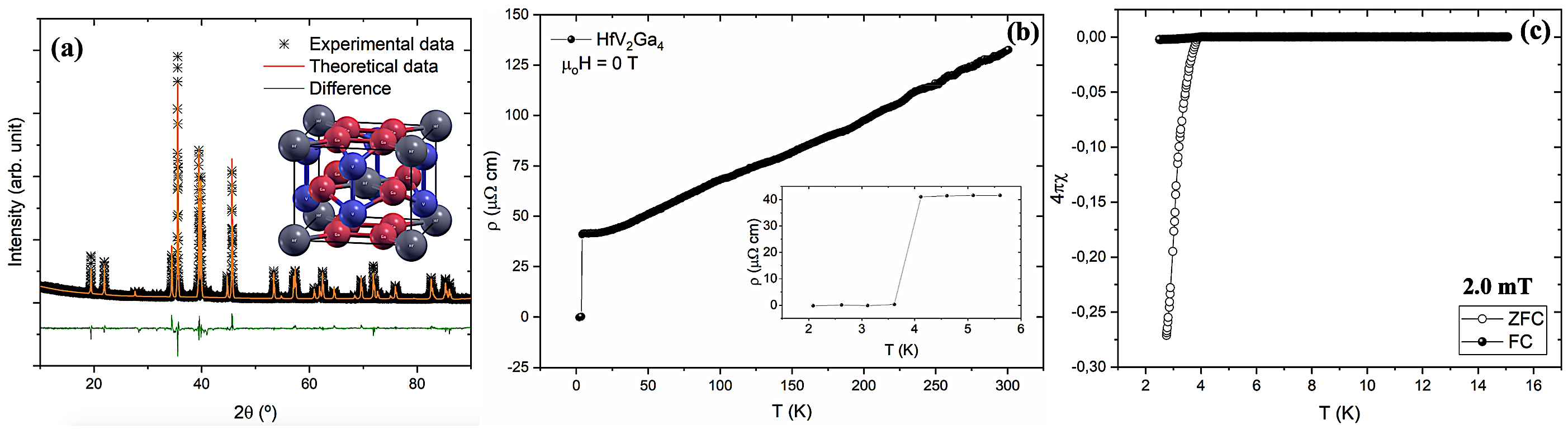}
\caption{(a) Represent the X-ray diffraction pattern (symbols) of HfV$_2$Ga$_4$ with Rietveld refinement (solid red line). Inset of (a) shows the tetragonal unit cell of HfV$_2$Ga$_4$, where Hf (gray), V (blue), and Ga (red) occupy at the 2$a$ (0, 0, 0), 4$d$ (0, $\frac{1}{2}$, $\frac{1}{4}$), and 8$h$ (0.303, 0.303, 0) Wyckoff positions, respectively. (b) Temperature variation of resistivity over the range 1.8 K $\le T \le$ 300 K, with the inset showing a drop in the resistivity at 3.9 K. (c) Temperature dependence of the magnetic susceptibility $\chi(T)$ in zero-field-cooled warming and field-cooled cooling (FCC) mode in presence of an applied magnetic field of 2 mT.}
\label{xrdmtrt}
\end{figure*}

\subsection{Crystal Structure, Resistivity, and Magnetization}

Fig.~\ref{xrdmtrt}(a) manifests a representative XRD pattern along with refinement which confirms the single-phase nature of HfV$_2$Ga$_4$ sample. It crystallizes in the YbMo$_2$Al$_4$ prototype tetragonal structure~\cite{Santos} with space group $I4/mmm$, no 139. The inset of Fig.~\ref{xrdmtrt}(a) displays a schematic illustration of tetragonal unit cell where Hf (gray), V (blue), and Ga (red) sites are at the 2$a$ (0, 0, 0), 4$d$ (0, $\frac{1}{2}$, $\frac{1}{4}$), and 8$h$ (0.303, 0.303, 0) Wyckoff positions, respectively. As shown in Fig.~\ref{xrdmtrt}(b), the $\rho(T)$ data exhibit a metallic behavior down to the superconducting transition at $T_{\mathrm{C}}$ = 3.9 K. As shown in Fig.~\ref{xrdmtrt}(c), magnetic susceptibility $\chi(T)$ data show $T_{\mathrm{C}}$ at 3.9 K, consistent with the value determined from $\rho(T)$ data.

\begin{table*}[htbp]
\caption{The parameters obtained from the fit to the
$\sigma_\mathrm{sc}(T)$ data of HfV$_2$Ga$_4$ using different gap models.}
\begin {tabular}{l l l l l l }
\hline       
\pbox{4cm}{Model} & g($\phi$)& \pbox{4cm} {Gap Value\\ $\Delta(0)$(meV)} & \pbox{4 cm}{Gap Ratio\\ 2$\Delta(0)$/k$_\mathrm{B}$T$_\mathrm{C}$} & \pbox{4cm} {Weight \\ factor} &~~ $\chi^{2}$ \\ \hline\\
$s+d$-wave & 1, $\cos(2\phi)$ & 0.98(1); 0.09(6) & 6.27(2); 1.14(1) & 0.57(1) & 1.3(2) \\
$s+s$-wave & 1& 1.08(3); 0.09(7) & 6.91(1); 0.57(5) & 0.87(2)& 2.2(3) \\ \\
\hline\\
anisotropy $s$-wave &$\frac{\vert 1+\cos(\phi) \vert}{2}$& 1.14(2) & 7.29(2) & 1.0 &2.3(1)\\
$s$-wave & 1 & 0.91(4) & 5.84(1) & 1.0 &2.3(2)\\
$d$- wave    & $\cos(2\phi)$ & 9.78(7) & 2.82(1) & 1.0 & 3.0(3) \\
     
\hline
\end{tabular}
\label{Tabel}
\end{table*}

\begin{figure*}[t]
\centering
\includegraphics[width=0.6\linewidth, height=0.5\linewidth]{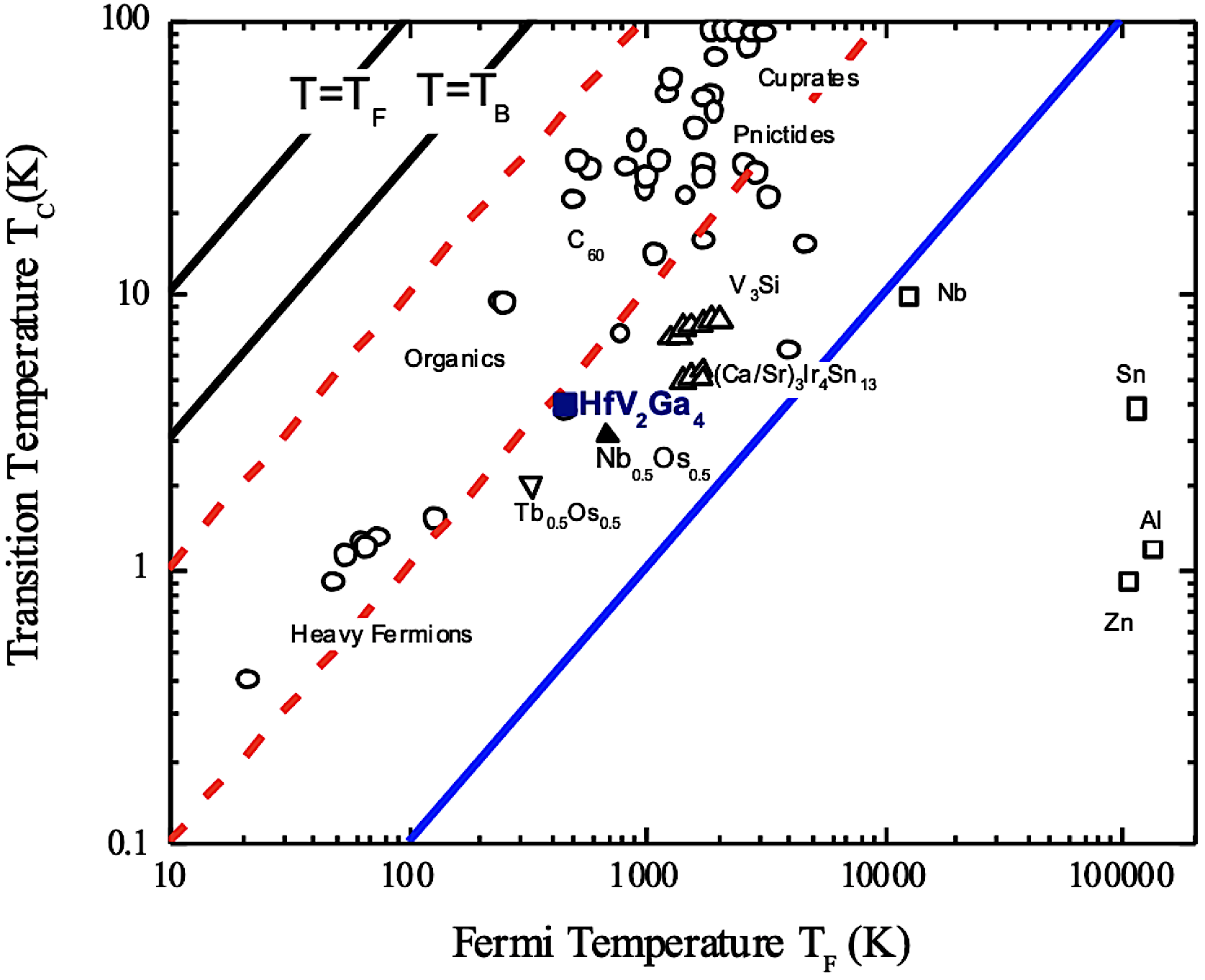}
\caption{(Color online) A schematic description of the Uemura plot. The exotic superconductors fall within a universal band for which 1/100$<T_{\mathrm{C}}$/$T_{\mathrm{F}}<$1/10, shown by the area between two red colors dashed lines in the figure. The solid black line resembles the Bose-Einstein condensation temperature ($T_{\mathrm{B}}$).~\cite{A1}}
\label{Uemura}
\end{figure*}

\subsection{TF-$\mu$SR:~Superconducting Parameters}

The muon spin depolarization rate observed below $T_{\mathrm{C}}$ is linked to the magnetic penetration depth. For a triangular\cite{EHB, Sonier,Chia,Amato} lattice $\frac{\sigma_\mathrm{sc}^2}{\gamma_\mathrm{\mu}^2}=\frac{0.00371 \times \phi_\mathrm{0}^{2}}{\lambda^4}$, where $\phi_{\mathrm{0}}$ = 2.07 $\times$10$^{-15}$ Tm$^{2}$, is the flux quantum number and $\gamma_\mathrm{{\mu}}/2\pi$ = 135.5 MHz T$^{-1}$, is the muon gyromagnetic ratio. Using London's theory \cite{Sonier} $\lambda_{\mathrm{L}}^2 = \frac{m^{*}c^{2}}{4\pi n_\mathrm{s}e^{2}}$, where $m^{*} = (1+\lambda_\mathrm{e-ph})m_\mathrm{e}$ is the effective mass and $n_\mathrm{s}$ is the density of superconducting carriers. $\lambda_{\mathrm{e-ph}}$ is the electron-photon coupling constant that estimated using McMillans relation \cite{McMillan,BhattacharyyaLaIr3, DasLaPt2Si2}:~$\lambda_\mathrm{e-ph} = \frac{1.04+\mu^{*}\ln(\Theta_\mathrm{D}/1.45T_\mathrm{C})}{(1-0.62\mu^{*})\ln(\Theta_\mathrm{D}/1.45T_\mathrm{C})-1.04}$, here $\mu^{*}$ is the repulsive screened Coulomb parameter with a typical value of $\mu^{*}$ = 0.12 and $\Theta_{\mathrm{D}}$ = 416.3 K, give $\lambda_{\mathrm{e-ph}} = 0.528(2)$. As HfV$_2$Ga$_4$ is a type II superconductor, supposing that approximately all the normal states carriers ($n_\mathrm{e}$) contribute to the superconductivity ($n_\mathrm{s} \approx n_\mathrm{e}$), the magnetic penetration depth $\lambda$, superconducting carrier density $n_\mathrm{s}$, the effective-mass enhancement $m^{*}$ have been estimated to be $\lambda_\mathrm{L}(0)$ = 797(4)~nm, $n_\mathrm{s} = 6.78(9) \times 10^{25}$ carriers m$^{-3}$, and $m^{*} = 1.528(2) m_\mathrm{e}$ respectively, for HfV$_2$Ga$_4$.

\subsection{Uemura Plot: Unconventional Superconductivity}

Uemura plot~\cite{U1} is crucial for classifying conventional and unconventional superconductors. In case of, Uemura classification scheme, which considers the correspondence between $T_\mathrm{C}$ and the effective Fermi temperature, $T_\mathrm{F}$, determined from TF-$\mu$SR measurement~\cite{A1}. Uemura {\it et al.} suggested that exotic superconductors, i.e. high $T_\mathrm{C}$ cuprates, heavy fermions, Chevrel phases and the organic superconductors, form a similar but different group, defined by a universal scaling of $T_\mathrm{C}$ with $T_\mathrm{F}$ such that 1/10$>(T_\mathrm{C}/T_\mathrm{F})>$1/100 as shown in Fig.~\ref{Uemura}. For conventional BCS superconductors 1/1000$>(T_\mathrm{C}/T_\mathrm{F}$). The estimated value of $T_\mathrm{C}/T_\mathrm{F}$ = 3.9/460.97 = 0.00846 for HfV$_2$Ga$_4$, which further suggest that HfV$_2$Ga$_4$ can be classified as an unconventional superconductor as shown in Fig.~\ref{Uemura}, according to Uemura's scheme ~\cite{U1}.

\begin{figure*}[t]
\centering
\includegraphics[width=\linewidth, height=0.5\linewidth]{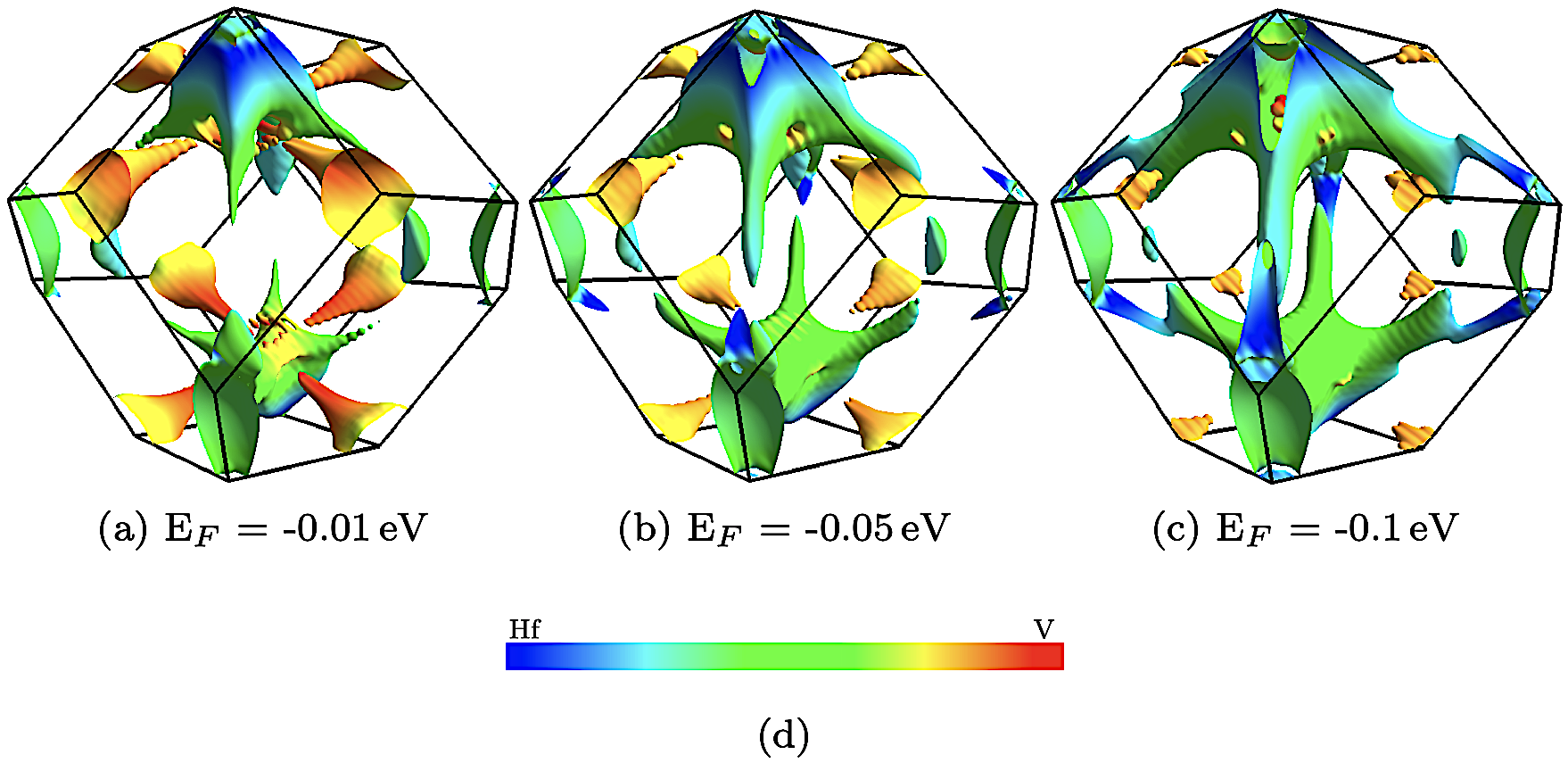}
\caption{Evolution of the orbital character projected over the Fermi surface, with hole doping according to the rigid band approximation.}
\label{fig:appendix}
\end{figure*}

\subsection{First-principles calculations:~Hole doping effect}

Fig.~\ref{fig:appendix}(a-c) shows modified Fermi surfaces by lowering the chemical potential in HfV$_2$Ga$_4$ by a few meV. The figures simulate, in the rigid band approximation, the depletion of electrons in going from the calculated HfV$_2$Ga$_4$ compound to HfV$_2$Ga$_{3.7}$, as experimentally measured. The hole \emph{jellyfish}- and \emph{horse-saddle}-pockets will continually increase, favoring spin-orbit coupling, nesting regions, and correlation effects, whereas the electron \emph{seashell}-pockets will gradually decrease. This hole doping can be reached, in a first approximation, utilizing Ga-deficiency control in HfV$_2$Ga$_4$ samples, considering the nature of Ga-Ga atomic bonding \cite{ferreira2018a}.

\end{document}